\documentclass[sn-mathphys-ay]{sn-jnl}

\usepackage{graphicx}%
\usepackage{multirow}%
\usepackage{amsmath,amssymb,amsfonts}%
\usepackage{amsthm}%
\usepackage{mathrsfs}%
\usepackage[title]{appendix}%
\usepackage{xcolor}%
\usepackage{textcomp}%
\usepackage{manyfoot}%
\usepackage{booktabs}%
\usepackage{algorithm}%
\usepackage{algorithmicx}%
\usepackage{algpseudocode}%
\usepackage{listings}%
\usepackage{textcomp}
\usepackage{setspace}
\usepackage{wrapfig}
\usepackage{subfigure}
\usepackage{wasysym}
\usepackage{derivative}

\theoremstyle{thmstyleone}%
\theoremstyle{thmstyletwo}%

\theoremstyle{thmstylethree}%

\begin{document}
\title[Article Title]{Lift generation on a sphere through asymmetric roughness using active surface morphing}


\author[1]{\fnm{Rodrigo} \sur{Vilumbrales-Garcia $\textsuperscript{\textdaggerdbl}$}}\email{rodrigga@umich.edu}
\author[2]{\fnm{Putu Brahmanda} \sur{Sudarsana$\textsuperscript{\textdaggerdbl}$}}\email{brahmsdr@umich.edu}
\author*[1,2]{\fnm{Anchal} \sur{Sareen}}\email{asareen@umich.edu}

\affil[1]{\orgdiv{Department of Naval Architecture and Marine Engineering}, \orgname{University of Michigan, Ann Arbor}, \postcode{48109}, \state{MI}, \country{USA}}

\affil[2]{\orgdiv{Department of Mechanical Engineering}, \orgname{University of Michigan, Ann Arbor}, \postcode{48109}, \state{MI}, \country{USA}}

\def\thefootnote{$\textsuperscript{\textdaggerdbl}$}\footnotetext{These authors contributed equally to this work}


\abstract{This study investigates a novel phenomenon of lift generation around a sphere by pneumatically manipulating its surface topology with an asymmetric roughness distribution. A comprehensive series of systematic experiments were conducted for Reynolds numbers ($Re = U_{\infty} d/\nu$, where $U_{\infty}$ is the fluid velocity, $d$ is the sphere diameter, and $\nu$ is the fluid kinematic viscosity) ranging from $6\times10^4$ to $1.3\times10^5$, and dimple depth ratios ($k/d$, where $k$ is the dimple depth) from $0$ to $1\times10^{-2}$ using a smart morphable sphere with one smooth and one dimpled side. The findings show that an asymmetrically rough sphere can generate lift forces up to 80\% of the drag, comparable to those produced by the Magnus Effect. The dimple depth ratio affects both the $Re$ at which lift generation begins and the maximum lift coefficient ($C_L$) achievable. The optimal $k/d$ for maximum lift varies with $Re$: deeper dimples are needed at low $Re$, while shallower dimples are more effective at high $Re$. For a fixed $Re$, increasing $k/d$ monotonically increases lift until a critical $k/d$ is reached, beyond which lift decreases. Particle Image Velocimetry (PIV) revealed that dimples delay flow separation on the rough side while the smooth side remains unchanged, resulting in asymmetric boundary layer separation, leading to wake deflection and lift generation. Beyond the critical $k/d$, the flow separation location moved upstream, increasing the size of the rear wake, reducing wake deflection, and thus decreasing lift. 
Overall, this study establishes the foundation for wake control over bluff bodies and paves the way for real-time manoeuvring applications.

} 





\maketitle
\section{Introduction}\label{sec1}

The study of flow around a sphere has been a subject of considerable interest for decades. The seminal study by \cite{Achenbach1972} revealed a sudden decrease in the drag coefficient ($C_D$) at a critical Reynolds number of $Re_c = 3.7 \times 10^5$, marking the transition from laminar to turbulent boundary layer and a significant delay in flow separation location—a phenomenon known as the drag crisis. \cite{Achenbach1974a} further reported that introducing roughness on the sphere shifts the critical Reynolds number ($Re_c$) at which the drag crisis occurs to a lower value. Higher roughness parameter lead to more perturbations, resulting in a lower $Re_c$. Dimples also have a similar effect, whereby they increase velocity fluctuations along the separating shear layer, enhancing momentum near the wall, delaying flow separation, and thereby reducing drag—a mechanism similar to the drag crisis \citep{Choi2006, Smith2010,Beratlis2019}. In addition, dimples have the benefit of not increasing drag considerably in the supercritical Reynolds number regime (the regime after the minimum in $C_D$ is reached), although that minimum value of $C_D$ is not as low as that of a smooth sphere.  \\

Additionally, the depth and distribution of dimples on a sphere can significantly affect the drag reduction. \cite{AOKI2012} found that altering dimple depth shifts the critical Reynolds number at which the drag crisis occurs. Deeper dimples reduce drag more effectively at lower Reynolds numbers but increase drag at higher Reynolds numbers compared to shallower dimples. While the drag reduction benefits of dimples are well known, they are highly dependent on incoming flow conditions. To address this limitation, \cite{vilumbrales2024} proposed an active surface morphing strategy that adjusts dimple depth on-demand or adaptively with changing flow conditions, thereby minimizing drag across a wide range of Reynolds numbers. They also found that the dimple depth ratio significantly influences the onset of the drag crisis and the minimum achievable drag. Using experimental data, they developed a model relating Reynolds number ($Re$) and dimple depth ratio ($k/d$) to optimize drag reduction across varying flow conditions. They implemented the control model and demonstrated real-time drag reduction of up to 50\% across the entire range of conditions examined.\\

While most studies focused on reducing drag using dimples, only a few have explored lift generation on dimpled spheres, primarily in the context of rotating golf balls \citep{Davies1949, beratlis2012, Lyu2020}. The lift generation in rotating golf balls is primarily due to the Magnus Effect, where the flow on the retreating side of the sphere separates later due to increased momentum, while the boundary layer on the advancing side separates earlier. This asymmetry in boundary layer separation causes an asymmetric pressure distribution on the surface, resulting in lift. This phenomenon is most commonly observed in the sports ball such as tennis, cricket, baseball, and golf ball \citep{mehta1985}. Conversely, the inverse Magnus Effect occurs at post-critical Reynolds numbers. Here, the boundary layer on the advancing side transitions to turbulence and separates later, while the boundary layer on the retreating side separates earlier, generating lift in the opposite direction to the Magnus Effect \citep{Davies1949, briggs1959, muto2012, kray2012magnus, beratlis2012, Kim2014, Lyu2020}. This highlights that asymmetric boundary layer separation on the two sides of the sphere is an effective strategy for generating lift.\\

Asymmetric boundary layer separation leading to lift generation can also be observed in sports, particularly in cricket. \cite{DeshpandeMittal2018} found that the seam of a cricket ball acts as a perturbation in the boundary layer flow. When positioned at certain seam angles (ranging from $10^\circ$ to $30^\circ$), the seam can create a laminar separation bubble that triggers a transition to turbulence on the seam side at a critical Reynolds number, while the boundary layer on the opposite side remains laminar, causing the ball to swing.  Furthermore, \cite{ShahMittal2023} showed that small spins of the cricket ball can induce a zigzag movement. This phenomenon occurs because altering the perturbation location along the ball can change the boundary layer flow, leading to fluctuating forces that cause the zigzag trajectory. These findings demonstrate that manipulating boundary layer separation through strategic perturbations, such as the seam on a cricket ball, could be an effective strategy for generating lift and trajectories of bluff bodies.\\

While several studies have explored lift generation on rotating spheres, both dimpled and non-dimpled, there is no systematic investigation, to the best of the authors' knowledge, on lift generation due to asymmetric roughness using active surface morphing. In this study, we address this research gap by performing a comprehensive series of systematic experiments and investigate the phenomenon of lift generation over a sphere by systematically varying the roughness parameter ($0 \leq k/d \leq 1\times10^{-2}$) using active surface morphing across a wide range of Reynolds numbers ($6\times10^4 \leq Re \leq 1.3\times10^5$). The specific research questions that will be explored are:
\vspace{0.05in}

\begin{itemize}
    \item Can we generate lift on a sphere solely through asymmetric roughness on its two sides? What's the maximum lift that can be produced?
 \item Can we manipulate the magnitude of the lift by changing the roughness parameter or the size of the roughness?
 \item How does the lift generated through this mechanism vary with Reynolds number?
 \item Is there an optimal roughness parameter that optimizes lift across a wide range of Reynolds numbers?
 \item How does the drag vary? Does increasing lift through this method also increase drag, as seen with the Magnus Effect?
\end{itemize}
\vspace{0.05in}
\textbf{Hypothesis:} Our hypothesis is that precise control of surface topography with asymmetric roughness can achieve asymmetric boundary layer separation on the two sides of the sphere, thereby generating lift. Additionally, controlling the roughness size will allow for precise control over lift generation. We also hypothesize that the optimal roughness size will be a function of the Reynolds number. This raises the question of whether we can control the roughness size on demand or adaptively with the Reynolds number to optimize lift generation. To investigate this, we will use a morphable sphere with the same working principle as reported in \cite{vilumbrales2024}. However, one side of the sphere will be smooth while the other side will be uniformly covered with dimples. This morphable sphere will allow us to precisely control the dimple depth on demand to study the systematic variation of the dimple depth (or roughness parameter) for a wide range of Reynolds numbers.

\section{Experimental Method}\label{sec2}
The experiments were performed in an open-circuit wind tunnel at the Department of Aerospace Engineering, University of Michigan, Ann Arbor with a test section 0.6 m wide, 0.6 m deep and 3 m long. The flow speed in the test section can be varied in the range $5\leq U_{\infty} \leq 25$~$\mathrm{ms^{-1}}$, which is measured using an inclined manometer and a pitot-tube manometer. The free-stream turbulence intensity levels were within $T_i=1.8\%$ for the range of flow speeds tested. Previous work by \cite{Achenbach1974a} has shown that $T_i<2\%$ have a limited impact on the drag production of a sphere in the sub-critical $Re$ range, which is investigated in this study. \\

\begin{figure}
	\begin{center}
	\includegraphics[width=14cm]{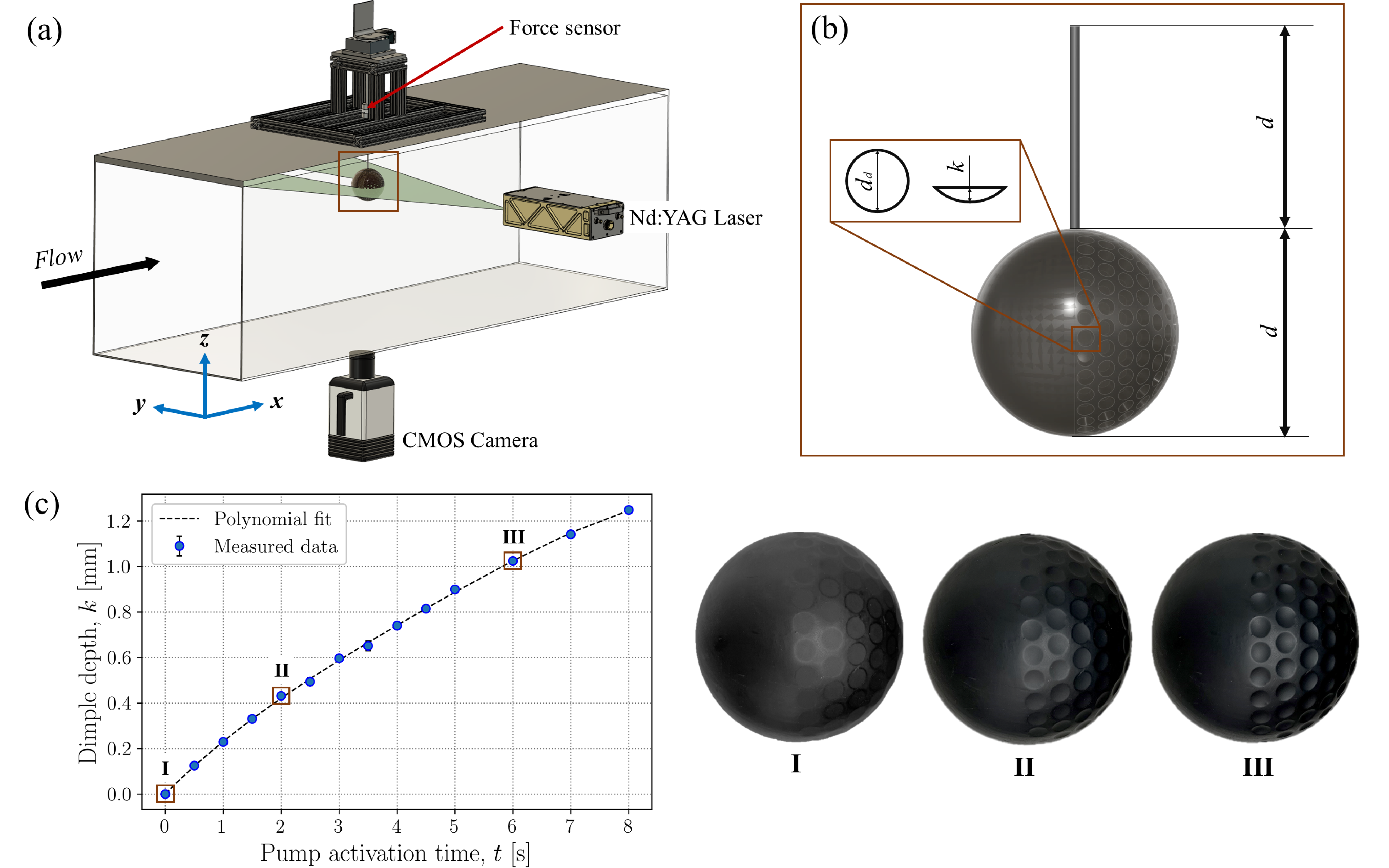}
	\caption{Details of the experimental methods, consisting of (a) A schematic of experimental setup, (b) Close-up view of the asymmetric morphable sphere, and (c) Morphable mechanism calibration showing the variation of dimple depth $k$ [mm] as a function of the pump activation time [s]. Three snapshots of the morphed asymmetric sphere are shown (right side) with the corresponding data points highlighted in figure (c).}
	\label{fig1}
	\end{center}
\end{figure}

A brief schematic of the experimental setup is shown in figure~\ref{fig1}(a). The setup is equipped with both a force measurement system and a 2D-2C Particle Image Velocimetry (PIV) system. The spherical model is 100 mm in diameter and is supported with a hollow cylindrical support rod of diameter 5 $\mathrm{mm}$. The exposed support rod length in the test section is equal to one sphere diameter. \cite{sareen2018vortex} reported that a diameter ratio (sphere diameter/support rod diameter) greater than 20 does not significantly influence the forces and wake behind the sphere. Additionally, they found that an exposed support rod length of one sphere diameter was sufficient to avoid any boundary effects. The blockage ratio, defined as the diameter of the sphere against the width of the test section ($d/W = 0.16$), is small enough to neglect its effect on the flow field \citep{Achenbach1974a}. More details of the experimental setup and validation is given in \cite{vilumbrales2024}.

The spherical model with asymmetric morphable dimples used in the current study is shown in figure~\ref{fig1}(b). The inner skeleton of the model is constructed using a stereolithography (SLA) resin 3D printer (Form 3, FormLabs) with a resolution of 25 $\mu$m. One hemisphere of the sphere is smooth, while the other one has holes uniformly distributed over the entire hemisphere. The dimpled configuration is dynamically matched with previous studies by \cite{Choi2006} and \cite{vilumbrales2024} with a dimple diameter to sphere diameter ratio of $d_d/d = 0.087$ and dimple area coverage ratio of $\phi = N_d d_d^2/(4d^2) = 55.6\%$, where $N_d$ is the total number of dimples ($N_d$ = 148) at the dimpled hemisphere. 

The inner spherical skeleton is enveloped with a pre-stretched thin latex membrane of thickness $0.276 \pm 0.001$ $\mathrm{mm}$ and shear modulus of $340 \pm 12$~$\mathrm{kPa}$. Dimples are actuated by depressurizing the sphere’s core using a vacuum pump (Kamoer KLVP6) with a maximum pressure of $-85$~$\mathrm{kPa}$, controlled by an Arduino UNO.
This setup allows a precise control of the dimple depth with a resolution of 0.01 mm by varying the pump’s activation time, shown in figure \ref{fig1}(c). Dimple depth is measured using a 3D laser scanner (scanCONTROL, Micro-Epsilon) with a 5~$\mathrm{\mu m}$ resolution, repeated for three times for each measurement. The deformation of the latex sheet due to flow is considered negligible since the maximum pressure at the highest $Re$ tested was estimated $\approx 0.3$~$\mathrm{kPa}$, which was an order of magnitude smaller than the vacuum pressure levels required to actuate the dimples. \\

The forces acting on the sphere were measured using an ATI mini40 IP65 six-axis force sensor, with a resolution of $0.01$ N in the streamwise and spanwise directions. The forces were acquired at a frequency of 1 kHz for 60 seconds, with each data point repeated three times. Drag coefficient, $C_D$, and lift coefficient, $C_L$, were obtained from the streamwise and spanwise forces respectively, evaluated as:

\begin{align*}
    C_D &= \frac{F_D}{(1/2)\rho A U_{\infty}^2};
    &
    C_L &= \frac{F_L}{(1/2)\rho A U_{\infty}^2}
\end{align*}

where, $F_L$ and $F_D$ are the lift force and drag force, respectively, $\rho$ is fluid density, $A$ is the projected area of the sphere ($A=\pi d^2 / 4$), and $U_{\infty}$ is the free-stream velocity. Following previous studies by \cite{sareen2018vortex, sareen2024}, we measured the drag of the isolated supporting rod with an end plate. This resulted in a constant drag coefficient of $C_D = 1.2$ across the entire tested $Re$ range, consistent with \cite{schlichting2016}. This value was then subtracted from the total drag recorded by the force sensor to isolate the drag of the sphere.

The flow field was captured using a 2D-2C Particle Image Velocimetry (PIV) measurement in an equatorial plane ($x-y$ plane) passing through the center of the sphere, as shown in figure~\ref{fig1}(a). The flow was seeded with polydisperse aerosol obtained by atomizing Di-Ethyl-Hexyl-Sebacat (DEHS) solution into particles of mean size 1~$\mathrm{\mu m}$ and density of $0.91$~$\mathrm{g/cm^3}$ using LaVision Aerosol Generator. An Evergreen 200 dual-pulsed laser with a pulse energy of 200 mJ was used to illuminate the particles. A Photron Nova R3-4K camera with a resolution of 4096 $\times$ 2304 pixels$^2$ was used to capture the PIV images. The camera was equipped with a 60 mm Nikon lens to provide a field of view of $400~\mathrm{mm} \times 250~\mathrm{mm}$. 1000 image pairs were captured for 67 seconds at a rate of 15 Hz. DaVis 11 software was used to cross-correlate the image pairs with an interrogation window size of $64\times64$ pixels with a 75\% overlap, providing a spatial resolution of $0.06d$.





\section{Results and Discussion}\label{sec3}

\begin{figure}
    \centering
    \includegraphics[width=0.95\textwidth]{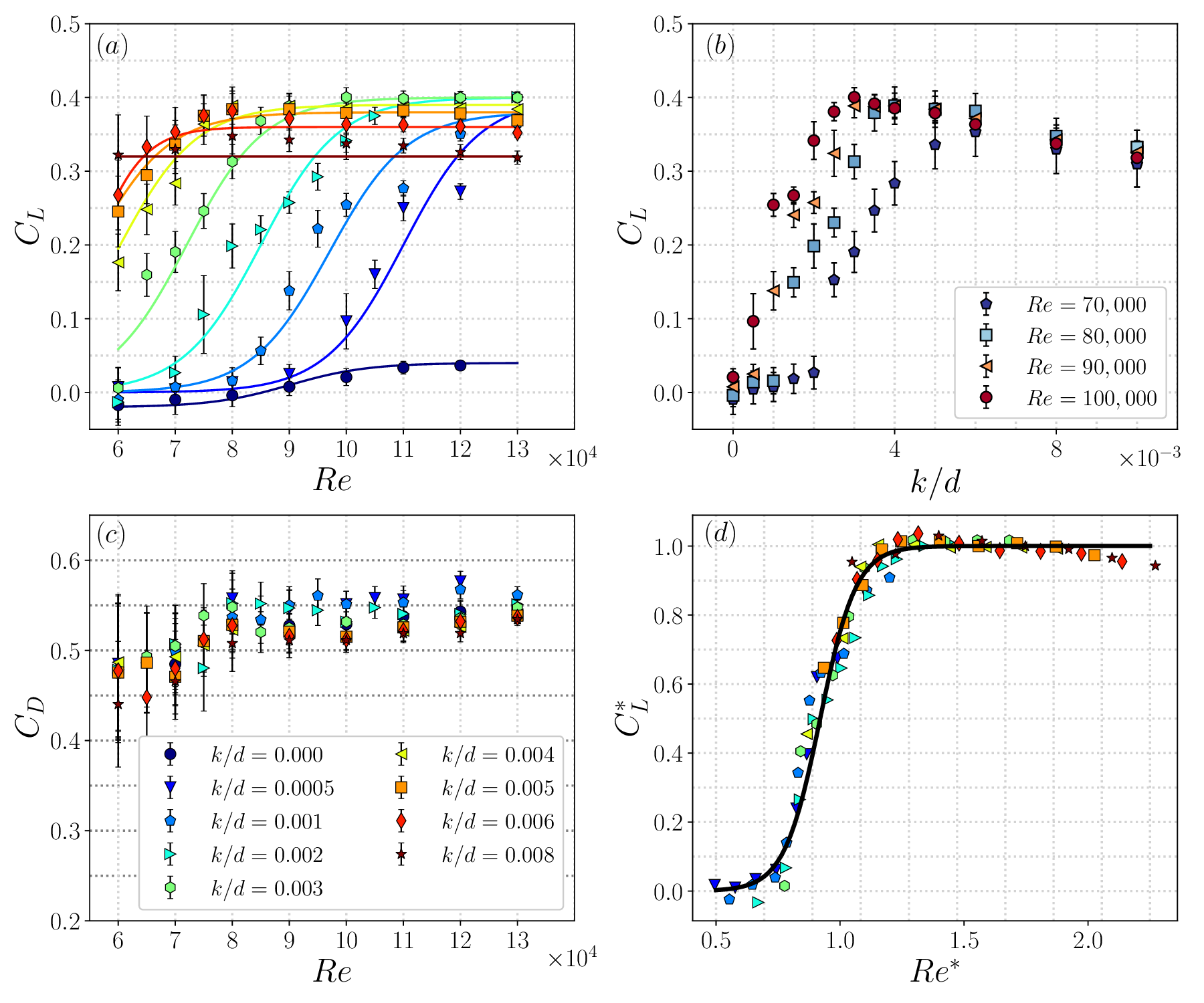}
    \caption{ (a) $C_L$ vs. $Re$ for varying $k/d$. (b) Evolution of $C_L$ with $k/d$ for various $Re$. (c) $C_D$ vs. $Re$ for various $k/d$, and (d) $C_L$ evolution for different $k/d$ presented in terms of effective $Re^*$ and normalised $C_L^*$. The legend corresponding to the data presented in (a), (c), and (d) is located at sub-figure (c).}
    \label{fig:fig2}
\end{figure}
\subsection{Lift and drag variations}
Figure \ref{fig:fig2} presents the evolution of the lift and drag coefficient ($C_L$, $C_D$) against the Reynolds number ($Re$) for several dimple depth ratios  ($k/d$) varying from  0 (corresponding to a smooth configuration) to $k/d=0.01$.
Three main observations can be noted in the forces analysis. First, the critical Reynolds number ($Re_c$) at which lift generation begins varies with the roughness parameter ($k/d$), decreasing as the dimple depth ratio increases. For example, the critical Reynolds number for $k/d=0.0005$ is
$Re_c=100, 000$, but decreases to $Re_c\approx 75, 000$ for $k/d=0.002$, as shown in figure~\ref{fig:fig2}(a). 
Second, the maximum achievable lift increases with the Reynolds number. As shown in figure \ref{fig:fig2}(b), at $Re=100, 000$, $C_{L_{max}}=0.4$ for $k/d=0.003$. Reducing the Reynolds number to $70, 000$ decreases the maximum $C_L$ generation by 10\%, resulting in a peak value of $C_{L_{max}}=0.35$ at a dimple depth ratio of $k/d=0.006$. In this scenario, the maximum lift force generated is approximately 80\% of the drag force at the same Reynolds number. The highest lift coefficient generated using this mechanism is comparable in magnitude to that produced by the Magnus effect, as reported by \cite{Kim2014}.

A third observation can be noted by analysing the evolution of $C_L$ at fixed $Re$ values, as displayed at figure \ref{fig:fig2}(b). Two regimes emerge for all considered Reynolds numbers. In the first phase, increasing dimple depth ratio at a fixed $Re$ value leads to a monotonic increase in the lift generation until maximum $C_L$ is achieved. In the second phase, further increases in $k/d$ begin to reduce lift generation. For example, at $Re=100, 000$, we observe a maximum lift coefficient of $C_L=0.4$. However, increasing the dimple depth ratio further decreases lift production by 25\%, resulting in a lift coefficient of $C_L=0.3$ at $k/d=0.01$ This indicates an optimal $k/d$ at a given $Re$ that maximizes lift.\\

Figure~\ref{fig:fig2}(c) shows variation of drag coefficient as a function of $Re$ for varying dimple depth ratios $k/d$. As evident from the figure, the drag coefficient of the sphere at $k/d=0$ remains constant at $C_D \approx 0.5$ in the subcritical $Re$ range,  consistent with prior benchmark studies by \cite{Achenbach1972, Choi2006}.
However, interestingly, the drag of the asymmetrically rough sphere remains almost constant as well for varying roughness parameter $k/d$. This is in contrast to the previous studies on Magnus Effect of a sphere \citep{kray2012magnus, Kim2014, sareen2024}, which found that the drag and lift are correlated. We will discuss this in more detail later in the following subsection.

Overall, the findings suggest that the evolution of $C_L$ at various $Re$ values is closely linked to the roughness parameter $k/d$. To further elucidate this finding, figure~\ref{fig:fig2}(d) presents the same data shown in figure~\ref{fig:fig2}(a), but in terms of an effective $Re^*$  and $C_L^*$. Our discussion of figure~\ref{fig:fig2}(a) highlighted that the $k/d$ parameter primarily affects two aspects of the lift evolution: the critical $Re$ and the maximum achievable $C_L$. To validate this, we develop empirical models relating these parameters to $k/d$ by extracting the $Re_c$ and $C_{L_{max}}$ information from figure~\ref{fig:fig2}(a) and fitting a curve through the data as:

\begin{equation}
Re_c(k/d)=75000 \times e^{(-431 \times k/d)}+ 49600,
\end{equation}
\begin{equation}
C_{L_{max}}(k/d)=-0.01 \times e^{(252 \times k/d)}+0.415.
\end{equation}

These models allow us to determine the critical $Re$ and maximum $C_L$ for any $k/d$ within the considered $Re$ range. By representing lift coefficient as $C_L^*=C_L/(C_{L_{max}}(k/d))$ against $Re^*=Re/(Re_c(k/d))$, we can collapse the $k/d$ evolutions onto one curve, as shown in figure~\ref{fig:fig2}(d). The data scaling based on $Re_c$ and maximum $C_L$ shows a logistic trend: an initial phase of negligible $C_L$ production, followed by a sharp increase and a stable phase at $C_{L_{max}}$. This nearly perfect collapse further supports the claim that the lift of an asymmetrically rough sphere can be predicted using only the roughness parameter $k/d$, within the considered $Re$. This simplified model provides a foundation for real-time closed-loop control applications. 


\subsection{Near wake measurements using Particle Image Velocimetry}

\begin{figure}
    \centering
    \includegraphics[width=1\textwidth]{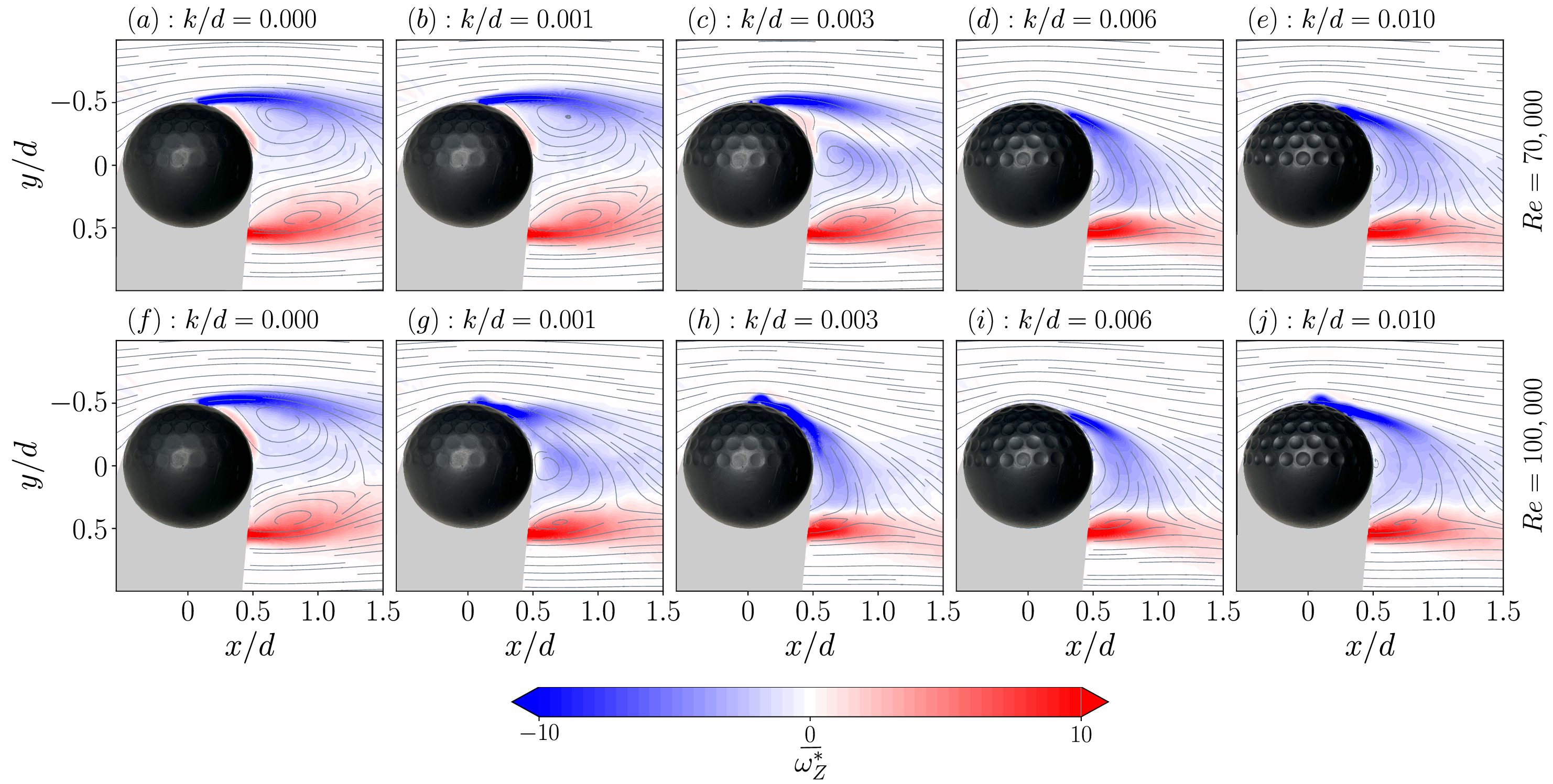}
    \caption{Time-averaged normalised streamwise vorticity $\overline{\omega_z}^*$ fields overlaid with time-averaged streamlines for several $k/d$ at (a)-(e) $Re=70, 000$ and (f)-(j) $Re=100, 000$. The gray region indicates the shadow of the sphere in the laser plane. Readers are referred to the supplementary video for real-time wake deflection with varying $k/d$ at a fixed Reynolds number of $Re = 70, 000$ and $Re = 100, 000$. }
    \label{fig:fig3}
\end{figure}

\begin{figure}
    \centering
    \includegraphics[width=0.9\textwidth]{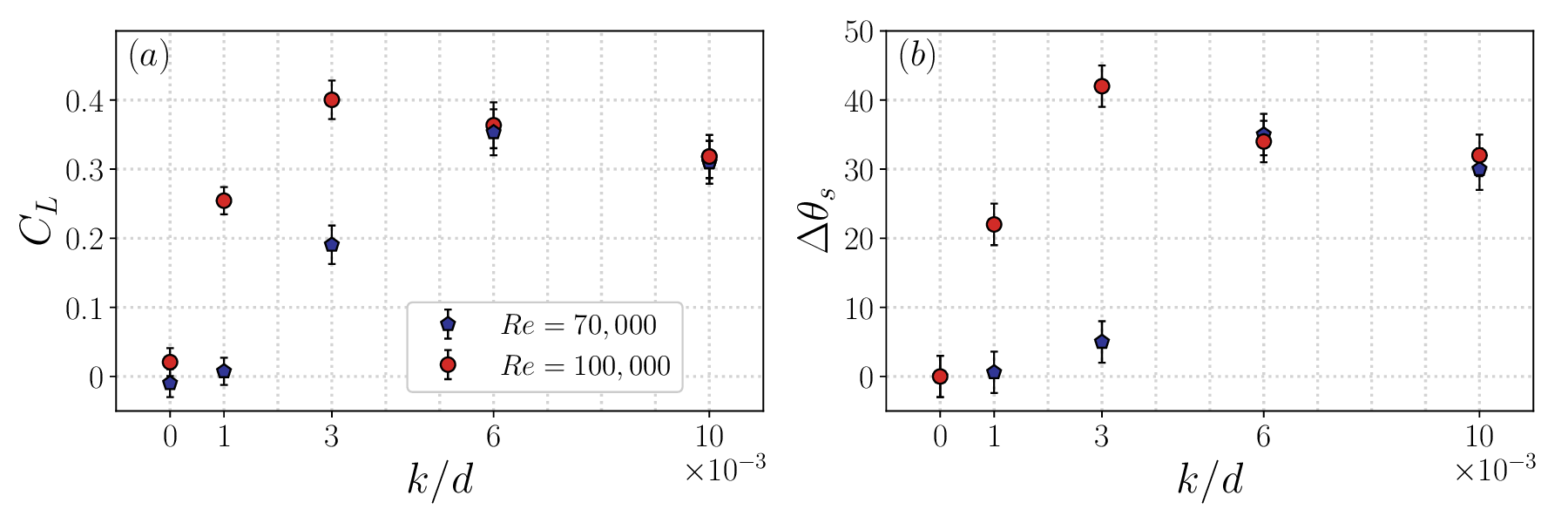}
    \caption{Variation of lift coefficient $C_L$ (a) and global separation angle $\Delta \theta_s$ (b) for varying $k/d$ at $Re=70, 000$ and $Re=100, 000$.}
    \label{fig:fig4}
\end{figure}

Figure \ref{fig:fig3} displays normalized vorticity fields, $\overline{\omega_z}^*=\omega_z d/U_\infty$, time-averaged over a $t^*=t U_\infty/ d$ of $6934$ at $Re=70, 000$ and $9909$ at $Re=100, 000$, corresponding to at least $1000$ vortex shedding cycles in both cases. The vorticity fields are overlaid with time-averaged streamlines for $k/d=0-0.010$ at a Reynolds number of $Re=70,000$ (top row) and $Re=100,000$ (bottom row). The correlated lift coefficient and the difference in global flow separation angle between the smooth and rough sides $\Delta \theta_s = \theta_{s_r}-\theta_{s_s}$ (measured from the leading stagnation point) on the surface of the sphere are also shown in figure~\ref{fig:fig4}. Here, subscripts $s$ and $r$ denote smooth and rough sides of the sphere. The separation angle on the sphere surface is estimated using MRS criterion \citep{Moore1958,Kim2014}: $U_\theta=0$ and $\partial U_\theta/ \partial r = 0$, where ($r$, $\theta$) are polar coordinates with the origin at the center of the sphere. The error in estimating the flow separation angle is $\pm 3^\circ$. This error is determined based on the distance corresponding to the window size used in the PIV processing ($64 \times 64$ pixels).

As evident from the vorticity plots, values of $k/d\leq 0.001$ do not significantly affect the wake characteristics at $Re=70, 000$. This is consistent with the variation of $C_L$ and $\Delta \theta_s$, as shown in figure~\ref{fig:fig4}, where $C_L$ and $\Delta \theta_s$ remain unchanged up to $k/d = 0.001$. However, as the roughness parameter is increased to $k/d=0.003$, the upper recirculation bubble moves downwards and closer to the sphere surface, breaking the symmetry of the wake, and hence generating lift, as shown in figure~\ref{fig:fig3}(c). This increase in lift correlates with a lift coefficient of $C_L =0.2$ (see figure~\ref{fig:fig4}(a)) and a separation angle difference of only 5 degrees (see figure~\ref{fig:fig4}(b)). On the other hand, for $k/d=0.006$, the near-wake is significantly deflected downward. This is correlated with a rapid rise in $C_L$ and $\Delta \theta_s$. This indicates that adding dimples to one side of the sphere delays the flow separation on that side, while the flow separation on the smooth side remains unchanged. This results in asymmetric flow separation between the two sides of the sphere, leading to a deflected wake and the generation of lift in the direction opposite to the wake deflection. 
This lift generation mechanism is similar to the Magnus Effect, where rotation causes wake deflection and produces lift in the opposite direction. In contrast, when $k/d$ is further increased to $0.01$, the wake deflection is reduced and the size of the rear wake region increases. This is associated with the flow separation location moving upstream, as evident from figure~\ref{fig:fig4}(b). This results in 15\% reduction in $C_L$ compared to the previous $k/d$ (see figure~\ref{fig:fig4}(a)). Readers are referred to the supplementary video that shows real-time wake deflection with varying $k/d$ at a fixed Reynolds number of $Re = 70, 000$ and $Re = 100, 000$.  

We observe a similar phenomenon at a much higher Reynolds number of $Re = 100,000$. Initially, the lift increases monotonically with the dimple depth ratio for $k/d\leq0.003$ due to delayed flow separation on the rough side. However, beyond a critical dimple depth ratio of $k/d = 0.003$, the flow separation location moves upstream, leading to a decrease in lift generation. Furthermore, we observe that the effects of the dimples are less sensitive to $Re$ for high dimple depth ratios, as shown by comparing figure \ref{fig:fig3} (e) and (j), and consistent with the variation of $C_L$ and $\Delta \theta_s$ in figure~\ref{fig:fig4}. Additionally, as $Re$ increases, the $k/d$ needed to reach maximum $C_L$ decreases, as seen by comparing $k/d=0.003$ for both $Re$. While at $Re=70, 000$ the wake exhibits characteristics similar to the smooth counterpart, a strong deflection is observed at the higher Reynolds numbers.

The present study finds that drag remains nearly constant with varying roughness parameters and Reynolds numbers. Previous studies on drag of dimpled spheres by \cite{Aoki2003} and \cite{Beratlis2019}, reported that global separation location alone cannot fully explain the drag behavior of a dimpled sphere. For instance, \cite{Aoki2003} observed that increasing dimple coverage delayed global separation but increased drag. Similarly, \cite{Beratlis2019} found through direct numerical simulations that dimples cause a local pressure penalty in the post-critical Reynolds number regime, accounting for up to 60\% of the total drag. These studies highlight the necessity of measuring flow within the dimples with high resolution to understand drag behavior fully. Such measurements are challenging to perform experimentally, underscoring the need for high-fidelity computations. Future research should focus on understanding drag variations with roughness parameter, as this question lies beyond the scope of the current paper.

\section{Summary and Conclusions}\label{sec4}

In this study, we conducted a series of systematic experiments to investigate lift generation using asymmetric roughness distribution on a sphere. By manipulating the surface topology with a smart morphable skin, we investigated the impact of varying dimple depth ratios ($0 \leq k/d \leq 1\times10^{-2}$) on lift across Reynolds numbers ranging from $6\times10^4$ to $1.3\times10^5$, through simultaneous measurement of forces and vorticity fields. 

Our findings reveal that asymmetric roughness can generate lift forces up to 80\% of the drag force, comparable to the Magnus effect. The critical Reynolds number ($Re_c$) at which lift production begins varies with the roughness parameter ($k/d$), decreasing as $k/d$ increases. Moreover, maximum lift increases with $Re$. For fixed Reynolds numbers, increasing $k/d$ initially increases lift coefficient ($C_L$) until reaching a critical $k/d$, beyond which $C_L$  declines, indicating an optimal $k/d$ that maximizes lift. This optimal dimple depth ratio varies with $Re$: deeper dimples enhance performance at low $Re$, while shallower dimples are more effective at high $Re$. 

The main mechanism underlying lift generation is asymmetric boundary layer separation induced by surface roughness. Roughness delays boundary layer separation on one side while the smooth side remains unaffected, causing wake deflection and lift generation. However, if the roughness parameter is increased beyond its optimal value, the separation location moves upstream, reducing lift production. To maximize lift production across Reynolds numbers, an effective strategy will involve adaptive morphing where dimple depth adjusts with flow conditions. In \cite{vilumbrales2024}, we demonstrated real-time control of dimple depth to achieve desired control across varying Reynolds number. Such a strategy could also be employed in the current study for optimizing lift across different Reynolds numbers. However, exploring this application is beyond the current study's scope and will be addressed in a future follow-up investigation.

This study underscores the potential of smart morphable skins to accurately control wake characteristics and generate lift on demand. This foundational research sets the stage for advanced adaptive flow control strategies, particularly in maneuvering and stability control for aerial and underwater vehicles. Future studies should focus on implementing real-time closed-loop control systems. Additionally, three-dimensional high-fidelity computations are recommended for deeper insights into these mechanisms. 

\section*{Acknowledgements}
The authors acknowledge the Center for Naval Research and Education (CNRE) funded by ONR at U-M for post-doctoral salary support for Vilumbrales-Garcia. Sudarsana acknowledges financial support from the Indonesia Endowment Fund for Education (LPDP) under the Ministry of Finance, Republic of Indonesia. The authors also acknowledge Parth Khokhani for his valuable discussions and contributions during the initial conceptualization phase of the project. We are grateful to the Department of Aerospace Engineering at U-M for providing access to the wind tunnel facilities.\\

\textbf{Declaration of Interests.}\\

The authors report no conflict of interest.

\bibliography{dimplepapers}

\end{document}